# MaterialsMap: A CALPHAD-Based Tool to Design Composition Pathways through feasibility map for Desired Dissimilar Materials, demonstrated with RSW Joining of Ag-Al-Cu

Hui Sun[a], Bo Pan[b], Zhening Yang[a], Adam M. Krajewski[a], Brandon Bocklund[c],

Shun-Li Shang[a], Jingjing Li[b], Allison M. Beese[a, d], and Zi-Kui Liu[a]

[a] Department of Materials Science and Engineering, The Pennsylvania State University,

University Park, PA 16802, USA

[b] Department of Industrial and Manufacturing Engineering, The Pennsylvania State University,

University Park, PA 16802, USA

[c] Materials Science Division, Lawrence Livermore National Laboratory, Livermore, CA 94550,

USA

[d] Department of Mechanical Engineering, The Pennsylvania State University,

University Park, PA 16802, USA

**Abstract:**

Assembly of dissimilar metals can be achieved by different methods, for example, casting, welding, and additive manufacturing (AM). However, undesired phases formed in liquid-phase assembling processes due to solute segregation during solidification diminish mechanical and other properties of the processed parts. In the present work, an open-source software named `MaterialsMap`, has been developed based on the CALculation of Phase Diagrams (CALPHAD) approach. The primary objective of `MaterialsMap` is to facilitate the design of an optimal composition pathway for assembling dissimilar alloys with liquid-phases based on the formation of desired and undesired phases along the pathway. In `MaterialsMap`, equilibrium thermodynamic calculations are used to predict equilibrium phases formed at slow cooling rate, while Scheil-Gulliver simulations are employed to predict non-equilibrium phases formed during rapid cooling. By combining these two simulations, `MaterialsMap` offers a thorough guide for understanding phase formation in various manufacturing processes, assisting users in making informed decisions during material selection and production. As a demonstration of this approach, a compositional pathway was designed from pure Al to pure Cu through Ag using `MaterialsMap`. The design was experimentally verified using resistance spot welding (RSW).

**Highlights**

- A CALPHAD-based open-source software tool, `MaterialsMap`, was developed to tailor the composition pathways to avoid undesired phases formed in joining dissimilar materials.
- A systematical approach is presented to design feasible path across materials based on equilibrium and nonequilibrium calculations of formation of phases .
- `MaterialsMap` incorporates three distinctive features to enhance the efficiency and accuracy of materials design: composition grids for different alloys, user-defined temperature ranges, and parallel computing.
- A feasible path to join Al and Cu with Ag as an interlayer was proposed by the feasibility map and validated through resistance spot welding experiments.

# 1. Introduction

Assembly of dissimilar metals has become increasingly crucial to create lightweight, high-performance, and multifunctional structures in various industries such as automotive [1–3], aerospace [4,5], marine [6], and biomedical [7]. However, the formation of undesired phases between dissimilar metals in liquid phase joining impairs targeted properties such as ductility and corrosion resistance [8–10] and may even result in cracks in the processed parts. Functionally graded materials (FGMs) are materials in which structures or compositions change with position [11]. FGMs are designed to mitigate the deleterious effects of an abrupt interface between two dissimilar materials [12]. Nevertheless, mixing different alloys during assembling dissimilar metals can result in the formation of brittle intermetallic phases, leading to cracking [13].

Several works have been done to computationally design dissimilar materials [14–19]. For example, Kirk et al. [14,15] introduced a computational framework based on equilibrium thermodynamic calculations in terms of the CALPHAD approach and a constraint satisfaction algorithm from the robotic motion planning community. However, in the joining process, a non-equilibrium heating and cooling process can lead to solute segregation in liquid phase during solidification, which is hard to capture through equilibrium thermodynamic calculations. Bocklund et al. [16,17] showed that the Scheil-Gulliver simulations [20,21] can predict the phases formed more accurately in the Ti-Invar functionally graded materials (FGM) than those relying on equilibrium calculations alone. Here, the Scheil-Gulliver simulations are used to predict phases formed under fast cooling conditions by assuming no diffusion in the solid, equilibrium at the solid/liquid interface, and infinite diffusion in the liquid. Moustafa et al. [18] proposed a Scheil-Gulliver ternary projection (STeP) diagram, which shows notable differences regarding phases formed in the Fe-Cr-Al system using `Thermo-Calc` [22]. In addition to the formation of phases, Yang et al. [19] successfully predicted hot cracking in FGMs from 316L Stainless Steel (SS316L) to Ni-20Cr (in wt. %), to Cr, to V, and to Ti-6Al-4V (in wt. %) based on the Scheil-Gulliver simulations using various hot cracking criteria in the literation.

The joining of Al and Cu is crucial for electronics packaging and the assembly of Li-ion battery cells due to the high conductivity and cost-effectiveness of Al-Cu joints [23,24]. However, three challenges remain: the disparate melting points (933 K for pure Al and 1357 K for pure Cu [25]), the different coefficients of thermal expansion ($23.6*10^{-6}$ $K^{-1}$ for pure Al and $16.5*10^{-6}$ $K^{-1}$ for pure Cu [26]), and the formation of brittle intermetallic compounds at the interface. Therefore, the design and production of reliable Al/Cu dissimilar joints remains a significant challenge in manufacturing [27,28].

In this study, a novel open-source `MaterialsMap` software is developed to design composition pathways between dissimilar materials through a feasibility map using high-throughput calculations at equilibrium combined with Scheil-Gulliver simulations. The term "feasible" indicates that the predicted mol fraction of deleterious phases is less than a defined threshold in equilibrium and Scheil-Gulliver simulations. The `MaterialsMap` incorporates equilibrium thermodynamic calculations and Scheil-Gulliver simulations to predict a range of phases formed under extreme slow and fast cooling conditions, respectively. Three unique functions are integrated in `MaterialsMap` to achieve efficient and accurate materials design. Firstly, a sampling grid of alloy compositions is employed in the `MaterialsMap`, allowing for the complex compositions of multicomponent alloys such as SS316, Inconel 625 (IN625), and Invar (Fe-36Ni in wt. %) as to serve as design components for FGMs, going beyond the typical case of pure elements. Secondly, the temperature ranges of the equilibrium calculation are automatically selected based on the composition-specific solidus temperature, enhancing the flexibility and accuracy of the equilibrium calculations. Thirdly, parallel computing is employed to accelerate the overall speed for both equilibrium calculations and Scheil-Gulliver simulations. Consequently, the `MaterialsMap` not only provides efficient and reliable high-throughput thermodynamic calculations, but also offers a systematic approach for the design of FGMs. As a demonstration, a compositional pathway has been proposed between Al and Cu via Ag and plotted through a feasibility map. The designed nonlinear compositio pathway is verified by resistance spot welding (RSW) experiments in the present study.

## 2. Methods and tools

### 2.1. Equilibrium calculations and Scheil-Gulliver simulations

Equilibrium thermodynamic calculations can be used to predict phases formed under slow cooling rates, while the Scheil-Gulliver simulations [20,21] can predict phases formed under fast cooling rates. The present work refrains from incorporating kinetic simulations involving nucleation and growth, given the intricate thermal dynamics inherent in FGM manufacturing, the requisite kinetic parameters, and the time-intensive nature of simulation processes. It's crucial to recognize that while `MaterialsMap` doesn't supplant kinetic simulations, it offers estimations of feasible composition pathways for FGMs, serving as a precursor to more resource-intensive kinetic simulations or experimental studies.

In the CALPAHD method, the Gibbs energy of individual phases is expressed as a function of temperature, pressure, and composition and covers both stable and unstable regions of phases [29–32]. Under constant temperature, pressure, and compositions thermodynamic equilibrium for a given system is obtained by the minimization of Gibbs energy of the system [33,34] with the amount of equilibrium phases as one of the outcomes [35]. As thermodynamic equilibrium represents an ideal state of a system, it is rarely reached during manufacturing, particularly during welding or AM. Nevertheless, equilibrium calculations are useful in informing the phases that may form in a system. In `MaterialsMap`, equilibrium calculations can be carried out for various conditions based on user needs through either the open-source software `pycalphad` [36,37], which is free and extensible, or the commercial software Thermo-Calc [22], which is the trusted industry standard.

The Scheil-Gulliver simulations [20,21] are based on three assumptions: (1) local equilibrium at the solid/liquid interface, (2) rapid diffusion in the liquid phase for perfect mixing, and (3) no diffusion in the solid phase. As the composition of the liquid changes during the solidification process, it becomes possible to predict solid phases that might not form in equilibrium calculations due to the significant change of the liquid composition. This capability reflects the dynamic evolution of the liquid phase's composition during the cooling process. In rapid cooling processes like welding and AM, diffusion in the liquid and Marangoni flow (i.e., the liquid transfer along the interface between liquid and solid due to surface tension) lead to mixing in liquid. For example, Bocklund et al. [16,17] adopted the Scheil-Gulliver simulations to study phases formed in the FGM between Ti-6Al-4V and Invar, showing that $Ni_3Ti$ predicted by Scheil was observed by electron backscatter diffraction (EBSD) but it was not present in equilibrium calculations. In

`MaterialsMap`, the Scheil-Gulliver simulations can be conducted using either of two available computational engines, i.e. `pycalphad` [36,37] and `Thermo-Calc` [22]. The simulations predict fractions and compositions of solid and liquid phases at various temperatures under each condition, efficiently delineating the phase boundaries formed in connection with rapid cooling.

**2.2. Computational tools: PyCalphad and Thermo-Calc**

Two software packages are employed in `MaterialsMap` as computational engines to conduct equilibrium calculations and Scheil-Gulliver simulations, i.e., `pycalphad`, and `Thermo-Calc`. PyCalphad [36,37] is a Python-based software for thermodynamic calculations using the CALPHAD method. Based on `pycalphad`, the Scheil-Gulliver algorithm has been implemented as an open-source Python package named `scheil` [16] for both the instantaneous and cumulative phase amounts across different temperatures. In the present work, `scheil` has been improved to distinguish disordered and partially/fully ordered phases in a partitioned order-disorder model [38,39], allowing the user to study order-disorder transition. Additionally, `scheil` has been implemented as a high-throughput process for systematically sampling grids in a multicomponent composition space in `MaterialsMap`.

`Thermo-Calc` [22] is a commercial Fortran-based software package that performs thermodynamic calculations and kinetic simulations including Scheil-Gulliver simulations. As a proprietary software, its efficient integration with open-source tools is challenging as reflected in the early development of our automated CALPHAD modeling efforts [40]. The `MaterialsMap` package developed in the present work can be connected to both `Thermo-Calc` and open-source `pycalphad` for high-throughput calculations and streamlined analysis of calculations results.

## 2.3. Experimental methods of Ag-Al-Cu for case study

To test `MaterialsMap`, a FGM for Al-Cu was designed and manufactured through RSW. An AA1000 Al alloy sheet (99.9% pure Al) measuring 40 mm × 15 mm × 0.5 mm and a C10100 Cu alloy sheet (99.9% pure Cu) with the same dimensions were chosen. As shown by the Al-Cu phase diagram in Figure 1, there are many intermetallic phases between Al and Cu, which are detrimental to the performance of the Al-Cu joints. To circumvent these phases, one or more intermediate alloys are needed. As both Al and Cu are fcc, the natural choice would be to search for fcc-based alloys, starting with pure elements with the fcc structure. Based on various Al-X and Cu-X binary systems with X being elements with the fcc structure, it is found that Ag is the only probable element. Within the Ag-Al binary phase diagram shown in Figure 1, there are two intermetallic solution phases forming from the liquid phase, and their structures are bcc and hcp, with large solubility ranges, thus significantly different from typical intermetallic compounds (IMCs) including those in the Al-Cu binary system shown in Figure 1. There is one additional phase forming at low temperature, also within a specific composition range. In the Ag-Cu binary phase diagram shown in Figure 1, there is no intermediate phase, and the positive interaction energy between Ag and Cu results in a miscibility gap between Ag-rich and Cu-rich fcc solid solutions.

It is thus plausible to prevent the formation of hard and brittle IMCs between Al and Cu by adding a Ag between them. In the present work, Al/Ag/Cu dissimilar RSW joints were created using a direct current welding machine (Milco Weld System, Milco Manufacturing in Warren, MI) with a 99.9% pure silver (Ag) interlayer measuring 15 mm × 15 mm × 0.2 mm. In this setup, the Al sheet was connected to the positive electrode, while the Cu sheet was connected to the negative electrode similar to the setup (sheets of Al alloy 1050 (AA1050) were connected to the positive electrode and T2 grade commercial pure Cu (CP-Cu) was connected to the negative electrode, a welding current of 18 - 23 kA was applied along with an electrode force of 1.5 kN. The entire welding process has a duration of 0.2 – 0.8 seconds.) reported in the literature [41]. Prior to the welding process, the Al sheet, Ag interlayer, and Cu sheet were carefully cleaned using abrasive paper and acetone to ensure proper bonding. During the RSW process, a welding current of 14 kA was applied along with an electrode force of 4 kN. The entire welding process has a duration of 0.8 seconds.

## 3. Architecture and capability of MaterialsMap

### 3.1. Architecture and Core functions of the MaterialsMap

`MaterialsMap` is a Python-based open-source software with MIT license hosted on Github [42] and available through the PyPI repository [43], making it easy to install using the "`pip install materialsmap`" one-line command. This command will install all the require dependencies except `Thermo-Calc`. `Thermo-Calc` needs to be installed separately because of the license. Its workflow is shown in Figure 2. With the input alloys being either pure elements or alloys, multiple compositions are generated based on the user's grid setting. The default for grid setting is 41, which means there are 41 divisions (42 points) along each axis and a total of 903 points for the full composition space. Both the equilibrium calculations and Scheil-Gulliver simulations are calculated for each composition using `Thermo-Calc` or `PyCalphad` with parallel computations. The calculation results are transferred to the JavaScript Object Notation (JSON) format for the construction of the feasibility map [19].

There are three feasibility maps, including the deleterious phases feasibility map with equilibrium calculations in a temperature range changed with solidus temperature at each composition, the deleterious phases feasibility map with a constant temperature range, and the crack susceptibility map for avoiding deleterious phases and five for avoiding crack in `MaterialsMap`, and additional customized ones can easily be added. When the goal is avoiding deleterious phases, the feasibility maps are typically equilibrium calculations in a temperature range changed with solidus temperature at each composition, equilibrium calculations at a constant temperature range, and Scheil-Gulliver simulations, with a default threshold of 10 mole % for the first two and 5 mole % for the third, which can be adjusted by users. A heatmap is used to illustrate the amounts of deleterious phases.

The five crack susceptibility criteria that have been implemented into the crack susceptibility map are: the Freezing Range (FR) [44], Crack Susceptibility Coefficient (CSC) [45], Kou Criteria [46], Improved Crack Susceptibility Coefficient (iCSC) [47], and Simplified Rappaz-Drezet-Gramaud (sRDG) [47], as detailed by Yang et al. [19].

In `MaterialsMap`, `pycalphad_eq()` function runs equilibrium calculations, and `pycalphad_scheil()` function runs Scheil-Gulliver simulations with a given starting temperature in `pycalphad`. However, if there are results from equilibrium calculations, the Scheil-Gulliver simulations will use the solidus temperature from equilibrium calculations to save time and accelerate the Scheil-Gulliver simulations. For using `Thermo-Calc`, `createEqScript()` generates macro files (i.e., the tcm files) for equilibrium calculations, and

`createScheilScript()` generates macro files (i.e., the tcm files) for Scheil-Gulliver simulations with a specified starting temperature. Using the generated tcm files, the users need to run them using `Thermo-Calc`. After completion, the functions `getEqdata()` and `getScheilSolidPhase()` will automatically gather data and save outputs using the JSON files. The `plotMaps()` function is used to visualize the results in JSON files for the eight feasibility maps mentioned above, with examples shown in Section 5.2 and Table S1 [19].

### 3.2. Unique capability in the MaterialsMap

Three unique capabilities are integrated into the `MaterialsMap` to improve efficiency and accuracy: a comprehensive sampling of the entire composition space through a grid based on the compositions of input alloys, an automated mechanism to choose the temperature range for equilibrium calculations based on the melting temperature, and parallel computing for high-throughput calculations.

A grid of compositions is created within the composition space according to the user-defined steps. For instance, in a ternary system such as Ag-Al-Cu, where there are 41 composition divisions along the x- and y-axes for the compositions of Al and Cu, respectively, there are 42 composition points on each axis, amounting to a total of 903 points. This comprehensive grid covers the entire composition plane for the Ag-Al-Cu system. Likewise, the composition space of a three-alloy system, denoted as Alloy1-Alloy2-Alloy3, can be generated using the same method as for the 3-element system. The visualization of the current feasibility map is restricted to 2-dimensional (2D) figures. In instances where the composition space is of higher dimensionality, such as in a four-alloy system, a series of 2D (3-component) feasibility maps can be sampled. In the next version of `MaterialsMap`, our recently developed `nimplex` high-performance library to compositional spaces will be used to more efficiently generate the composition grids, particularly for high dimensional space, and rapidly calculate corresponding gradient maps [48].

As discussed in Section 1, the isothermal phase diagrams from equilibrium calculations have been extensively used in designing composition pathways for fabricating FGM with tailored phases and properties. Nevertheless, maintaining a constant temperature range for equilibrium calculations may not always be appropriate for predicting phase formation across a broad composition range, especially in cases where there is a significant difference in melting temperatures, such as the 425K disparity between Al and Cu in the Al-Cu system [49] as shown in Figure 1. To address this issue, `MaterialsMap` allows the user to define a range of temperatures for equilibrium calculations. Two-thirds of the solidus temperatures at each composition are identified as the diffusion freezing temperature [50]. Below the diffusion freezing temperatures, it is assumed that there are no solid phase transformations due to the sluggish kinetics. In the present package, the user has the flexibility to choose any temperature range or use a variable temperature range between the solidus and 2/3 of the solidus for equilibrium calculations, with the latter being the default setting. Different outcomes of these choices for temperatures for equilibrium calculations are discussed in Section 5.2. This feature increases the flexibility and accuracy of the equilibrium calculations.

`MaterialsMap` employs the Python multiprocessing pool library to accelerate equilibrium calculations and Scheil-Gulliver simulations using multicore processors in distributed computing environments. The `pycalphad_eq()` and `pycalphad_scheil()` functions are designed to be use `Pool` library for seamless integration into parallel computing. Enhancements have been made to the `Pool` package to incorporate features that enable the display of time and process information throughout the parallel computing process, facilitating better monitoring and management of computational tasks.

## 4. Steps to run MaterialsMap

In this section, a step-by-step tutorial for `MaterialsMap` is presented. All the functions employed in this section have been introduced and discussed in Section 3.2. Comprehensive details of each step can be found in Table S1 in supplementary materials. A demonstration for the Ag-Al-Cu system is presented at Section 5. `MaterialsMap` code can be installed from PyPI using "`pip install materialsmap`" and or from source hosted at GitHub (materialsmap.phaseslab.org). Steps to run `MaterialsMap` are as follows:

1. This The input file for `MaterialsMap` is the TDB file, and serval settings are needed including three alloys (or elements) of interest, composition unit (`eleAmountType`), temperature range for equilibrium calculations, pressure, Scheil-Gulliver calculations, gird density to sample the composition space (`ngridpts`), and the number of simulations in each TCM files (`maxNumSim`). All those settings are detailed in the Table. S1 in supplementary materials. With these inputs and settings, `MaterialsMap` generates two outputs: `setting.npy` to record all the inputs and settings for user reference and `composition.xlsx` for the compositions obtained from the sampling process.

2. With `setting.npy` and `composition.xlsx` files, `pycalphad_eq()` and `pycalphad_scheil()` run equilibrium and Scheil-Gulliver calculations using `pycalphad`, and save the results in a JSON file. `createEqScript()` and `createScheilScript()` conduct the equilibrium and Scheil-Gulliver calculations through `Thermo-Calc`, and the resulted exp files are processed by `getEqdata()` and `getScheilSolidPhase()` to generate a JSON file.

3. Utilizing the JSON files (either from `pycalphad` or `Thermo-Calc`), `plotMap()` will generate three types of deleterious phases feasibility map with equilibrium calculations in a temperature range changed with solidus temperature at each composition, deleterious phases feasibility map with a constant temperature range, and the crack susceptibility map as described in Section 3.2 and demonstrated in Section 5.2.

## 5. Demonstration of the MaterialsMap code for the Ag-Al-Cu system

The Ag-Al-Cu system is chosen to demonstrate the application of `MaterialsMap` to design a feasible path from pure Al to pure Cu by avoiding the formation of IMCs between Al and Cu. The Ag-Al-Cu system was modeled by Witusiewicz et al. [49] using the CALPHAD method. The designed composition pathway was tested experimentally by the RSW process as detailed in Supplemental Material. There are 12 phases in the Ag-Al-Cu system, i.e., four solutions phases including liquid, FCC_A1 (space group of $Fm\bar{3}m$), BCC_A2 (space group of $Im\bar{3}m$), and HCP_A3 (space group of $P6_3/mmc$) and 8 non-stoichiometric IMCs from binaries including CUB_A13 (15 – 23 at. % Al in Ag-Al), ETA (50 – 51 at. % Cu in Al-Cu), DELTA (60 at.% Cu in Al-Cu), THETA (33 at.% Cu in Al-Cu), EPSILON (54 – 57 at.% Cu in Al-Cu), ZETA (55 at. % Cu in Al-Cu), GAMMA_D83 (63 – 69 at. % Cu in Al-Cu), and GAMMA_H (66 – 69 at. % Cu in Al-Cu). 8 non-stoichiometric IMCs are modeled with all three elements.

## 5.1. Predictions of feasibility maps by MaterialsMap

To facilitate the demonstration, a Jupyter Notebook is developed with all details provided as Supplementary Material with `ngridpts=41`, weight fraction as composition unit, default number of simulations in TCM files, the `database` path linked to the TDB file, and the `pressure` and `temperature` of 101325 Pa and 600 – 2000 K, respectively.

Figure 3 shows the predicted feasibility map of Ag-Al-Cu with the temperature range of equilibrium calculations as a function of compositions. In the present work, the thresholds of deleterious phases are set as default values as discussed in Section 3.1. In Figure 3, the Equilibrium and Scheil infeasible region (purple color) indicates that both equilibrium calculations and Scheil-Gulliver simulations predict the presence of deleterious phases exceeding the thresholds. The Equilibrium infeasible, Scheil feasible region (red color) signifies that equilibrium calculations predict the presence of deleterious phases exceeding the threshold, whereas Scheil-Gulliver simulations do not. The Equilibrium feasible, Scheil infeasible region (blue color) indicates that Scheil-Gulliver simulations predict the presence of deleterious phases exceeding the threshold, while equilibrium calculations do not. The Equilibrium and Scheil feasible region (green color) represents situations where neither equilibrium calculations nor Scheil-Gulliver simulations predict the presence of deleterious phases exceeding the thresholds. As noted above, the temperature range of equilibrium calculations is between solidus and its 2/3 for each composition. From the feasibility map in Figure 3, the feasible path to join Al-Cu without forming IMCs (avoid purple, blue and red areas) is from Al via Ag to Cu as shown by the black lines.

To further investigate the effect of temperature range of equilibrium calculations, the feasibility map is also calculated with a fixed temperature range of 600 K – 1500 K for all compositions with the low temperature being about 2/3 of the melting temperature of Al (993 K) and plotted in Figure 4 with the same color scheme used in Figure 3. A larger equilibrium infeasible region from 2Al-0Cu-98Ag to 8Al-0Cu-92Ag at the Ag rich corner is observed in Figure 4 than that in Figure 3. The fixed temperature range thus tends to overestimate the infeasible region than the variable temperature ranges with local compositions due to the low temperature of 600 K being 0.44 of the melting temperature of Cu at 1357 K. The wider temperature range in high Cu concentration regions results in more IMCs as shown in Figure 5 for a given composition of 4.5Al-2Cu-93.5Ag (in wt.%) in terms of equilibrium calculations and Scheil-Gulliver simulations. For example, the CUB_A13 phase appears in equilibrium calculations from 600 – 643 K in Figure 5 (a) with 643 K being 467 K lower than the solidus temperature of 1110 K at 4.5Al-2Cu-93.5Ag (in wt.%), thus wider infeasible region.

Figure 6 shows the heatmap of equilibrium calculations at 600 K with the amounts of six deleterious IMCs in red color and three solutions phase in black color, total nine phases. All IMCs form in the middle of the Al-Cu-Ag phase diagram with phase fractions increasing with the Cu content, except CUB_A13 forms in the Ag-rich region with its fraction decreasing with increasing Cu content. Figure 7 shows the predicted heatmap by Scheil-Gulliver simulations, with seven deleterious phases in red color and three solutions phases in black color, a total of 10 phases. The fraction of DELTA is less than 0.03 and not plotted. Solution phases are usually ductile, but the mismatch of their properties, such as the coefficient of thermal expansion (CTE), may result in the failure of the Al/Cu joint due to the formation of cracks during processing. Therefore, the feasibility map is fundamental for future analysis of mismatches of different properties based on the formation of different phases. It is noted that brittle IMCs form in broader composition ranges than those in Figure 6.

Beyond phases predicted by `MaterialsMap`, it is easy to calculate various phase-related material properties like CTE and Young's modulus. For example, Yang et al. [19] predicted the hot cracking feasibility map using the change of liquid phase fraction with respect to that of solid phases, agreeing well with experimental measurements of FGM from 304L Stainless Steel (SS304L) to IN625.

### 5.2. Experimental validations of the phases formed in Al-Ag-Cu

Figure 8 (a) shows the optical image of successful built of join of Al-Cu with Ag as interlayer, demonstrating the effectiveness of the path designed by `MaterialsMap` without cracking. Figure 8 (b, c) displays the EDS images of the sample, providing the distribution of elements in the built samples with 100 at. % Cu initially, 80 at. % Ag and 20% Cu at approximately 83 μm, 65 at. % Al, 30 at. % Ag, and 10 at. % Cu at 205 μm, and 100 at. % Al from 500 to 588 μm, in agreement with the designed pathway shown as the black line in Figure 3.

XRD in Figure 8 (d, e) indicates the formation of both $Ag_2Al$ (HCP_A3) and $Ag_3Al$ (CUB_A13 at the fusion zone of the Al-Ag-Cu RSW joint. However, distinguishing between $Ag_2Al$ (HCP_A3) and $Ag_3Al$ (CUB_A13) is challenging due to their similar XRD peak patterns. EBSD in Figure 8 (g, f) confirms that only $Ag_2Al$ (HCP_A3) is formed during the welding processing based on Kikuchi patterns, in agreement with the predictions shown in Figure 7.

The experimental observations show that there are no detrimental IMCs formed in the build, validating the designed compositional path from Al to Cu with Ag as an interlayer from Figure 3. The observed formation of Ag$_2$Al (HCP_A3) is predicted by both the Scheil-Gulliver simulations in Figure 7 and the equilibrium calculations in Figure 6. However, the Scheil-Gulliver simulations also predict the BCC_A2 phase will form between Ag and Cu, but it was not observed by experiments. This is probably due to the fact that the BCC_A2 phase forms above 880 K and transforms to Ag$_2$Al (HCP_A3) below 880 K as shown in Figure S1. The Scheil-Gulliver simulations cannot predict solid-to-solid phase transformations, so cannot capture the phase transition from BCC_A2 to HCP_A3. On the other side, equilibrium calculations predict the formation of CUB_A13, which was also not observed by experiments. This may be due to sluggish kinetics since CUB_A13 only forms below 728 K as shown in Figure S1.

## 6. Conclusions

In the present work, a Python-based open-source software, `MaterialsMap`, has been developed for designing optimal pathways for joining dissimilar materials based on high-throughput equilibrium calculations and Scheil-Gulliver simulations and released to the public via Github [42] and PyPI [43]. The Al-Ag-Cu system is employed as an example to demonstrate the predictions of feasibility maps generated by `MaterialsMap`. Experimental validation indicates the Al-Ag-Cu joined without forming detrimental IMCs as predicted by the `MaterialsMap`. The present work shows that:

- `MaterialsMap` incorporates three unique features: composition grids of different alloys, user-defined temperature range, and parallel computing. These features broaden the applications of thermodynamic calculations and enhance the efficiency and accuracy of the materials design.

- Feasibility maps, considering the presence of deleterious phases and crack susceptibility during liquid phase joining, can be efficiently predicted from `MaterialsMap`, which provides information on both the equilibrium and non-equilibrium phases, enabling a systematic approach to design FGMs through high-throughput thermodynamic calculations.

- The predicted feasibility map reveals a feasible composition pathway to join Al and Cu using Ag as an interlayer without the formation of IMCs, which is validated experimentally in the present work. This example reinforces the applicability and reliability of feasibility map predictions.

## Acknowledgments

The authors acknowledge financial support by the Office of Naval Research (ONR) under Contract No. N00014-21-1-2608 and National Science Foundation (NSF) via Award Nos. CMMI-2050069, CMMI-2226976, and FAIN-2229690. Simulations were performed on the Roar supercomputer at the Pennsylvania State University's Institute for Computational and Data Sciences (ICDS). Part of the present work was performed under the auspices of the U.S. DOE at Lawrence Livermore National Laboratory under Contract No. DE-AC52–07NA27344.

**Figures:**

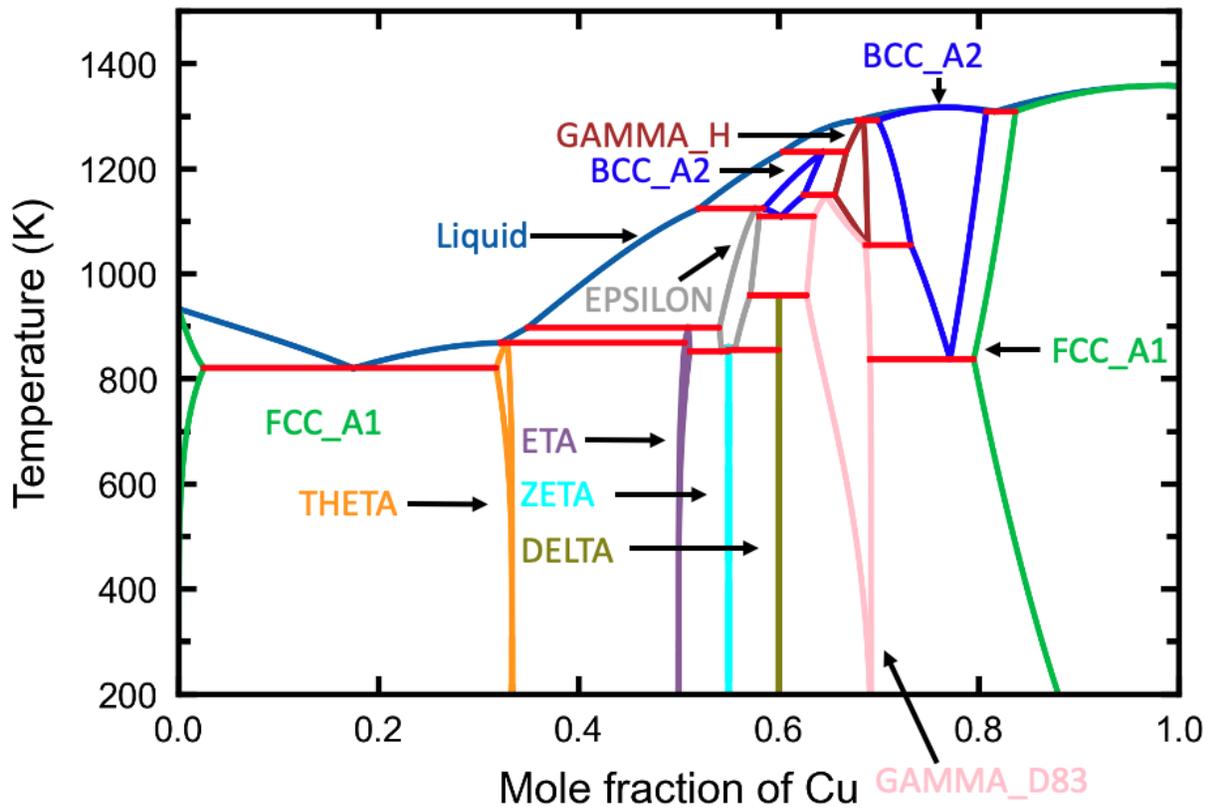

Figure 1 Predicted Al-Cu phase diagram based on the CALPHAD modeling by Witusiewicz et al. [49].

Figure 2 Workflow of MaterialsMap.

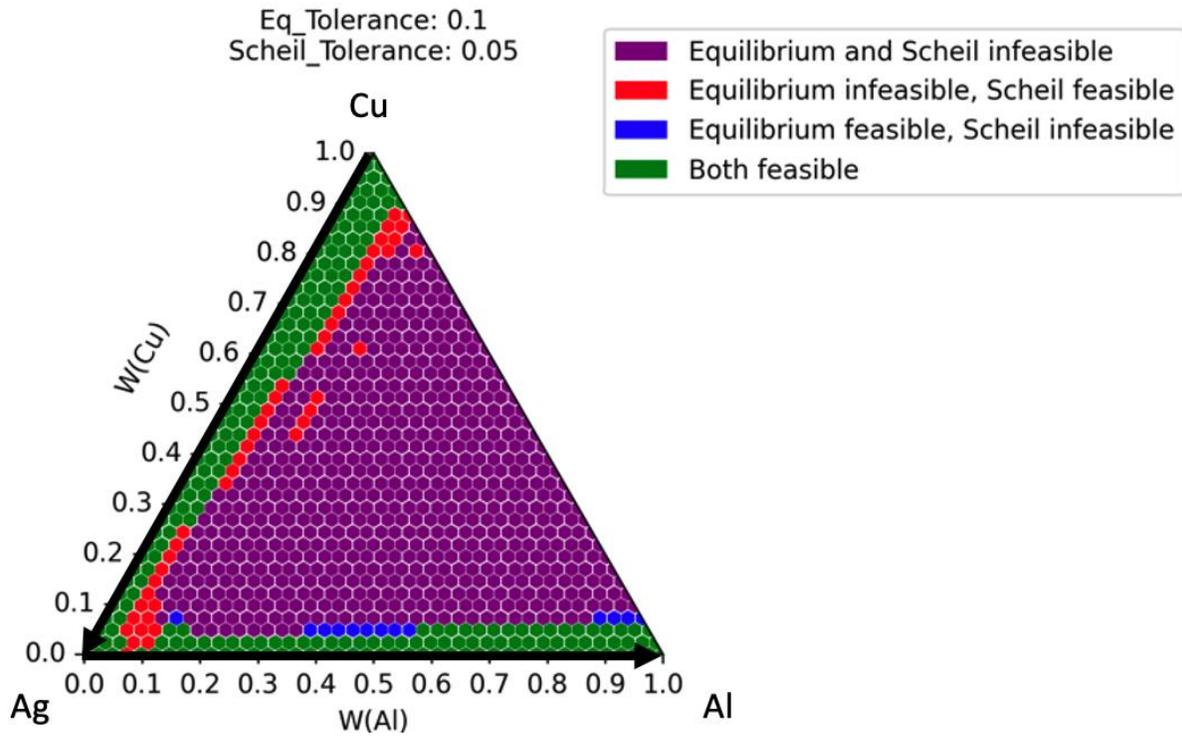

Figure 3 Predicted feasibility map with equilibrium temperature as 2/3 of solidus at various compositions and the black line denoting the designed composition pathway to join Al-Cu.

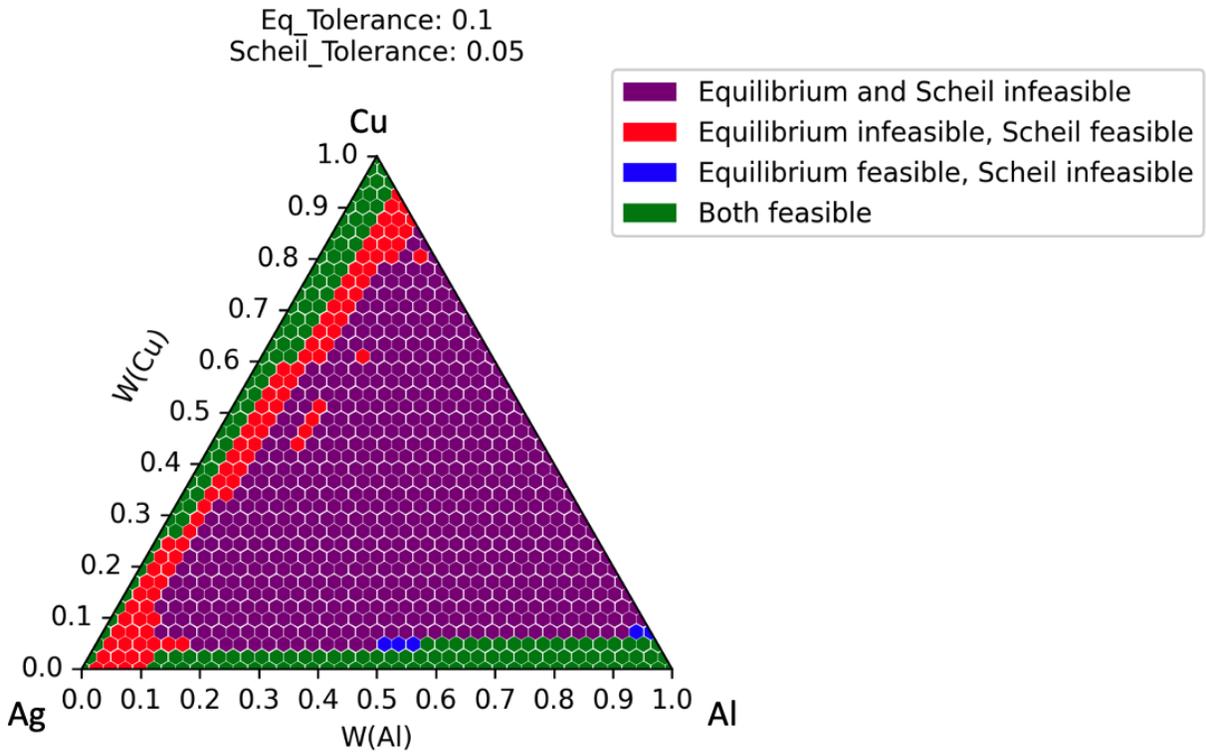

Figure 4 Predicted feasibility map with the temperature range from 600 to 1500 K for equilibrium calculations.

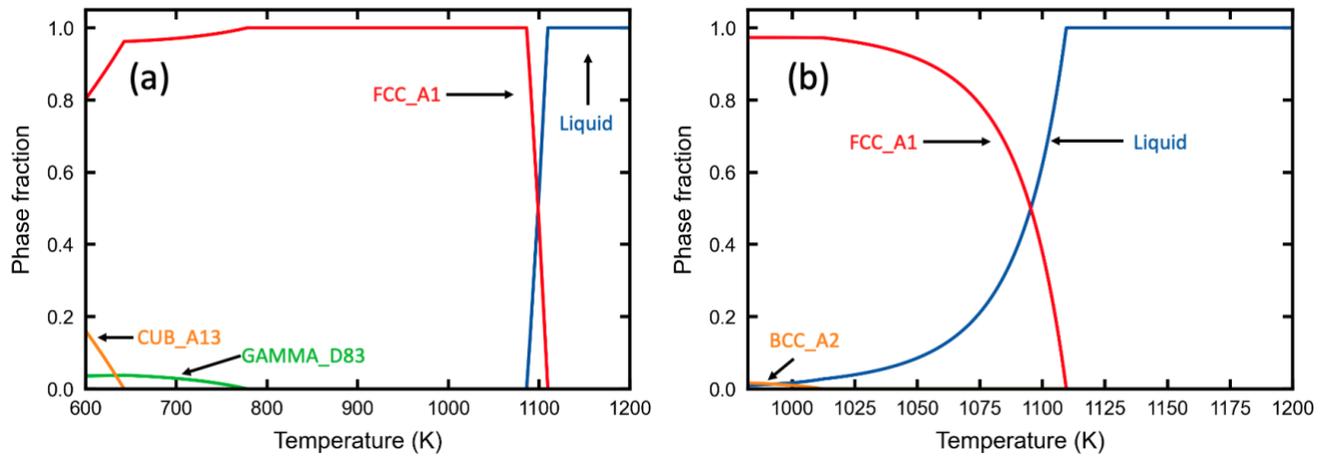

Figure 5 (a) Equilibrium and (b) Scheil-Gulliver simulations [20,21] at the composition of 4.5 wt. % Al, 2 wt. % Cu, and 93.5 wt. % Ag from `pycalphad`.

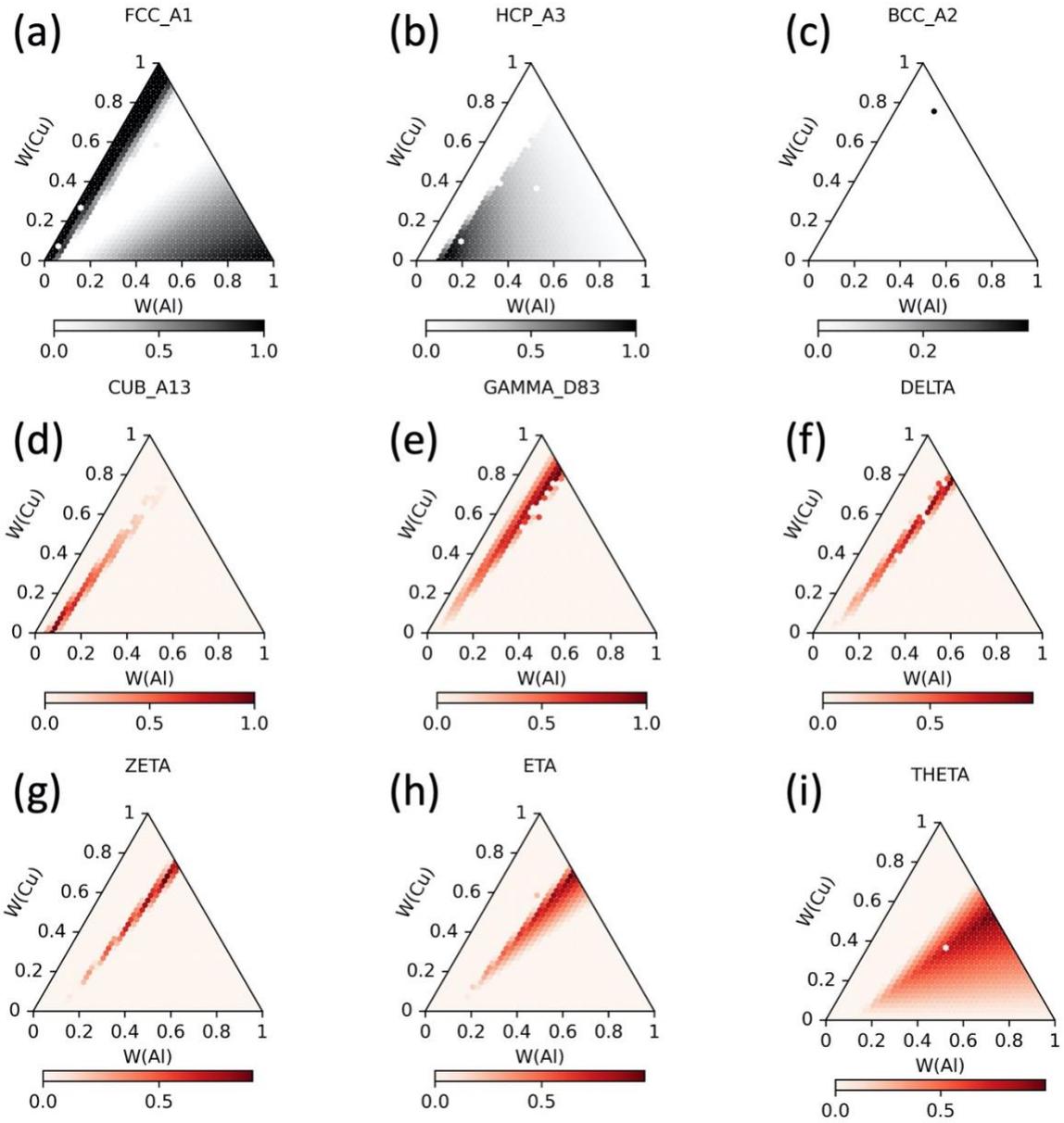

Figure 6 Heatmaps of predicted equilibrium phase fractions in Ag-Al-Cu at 600 K.

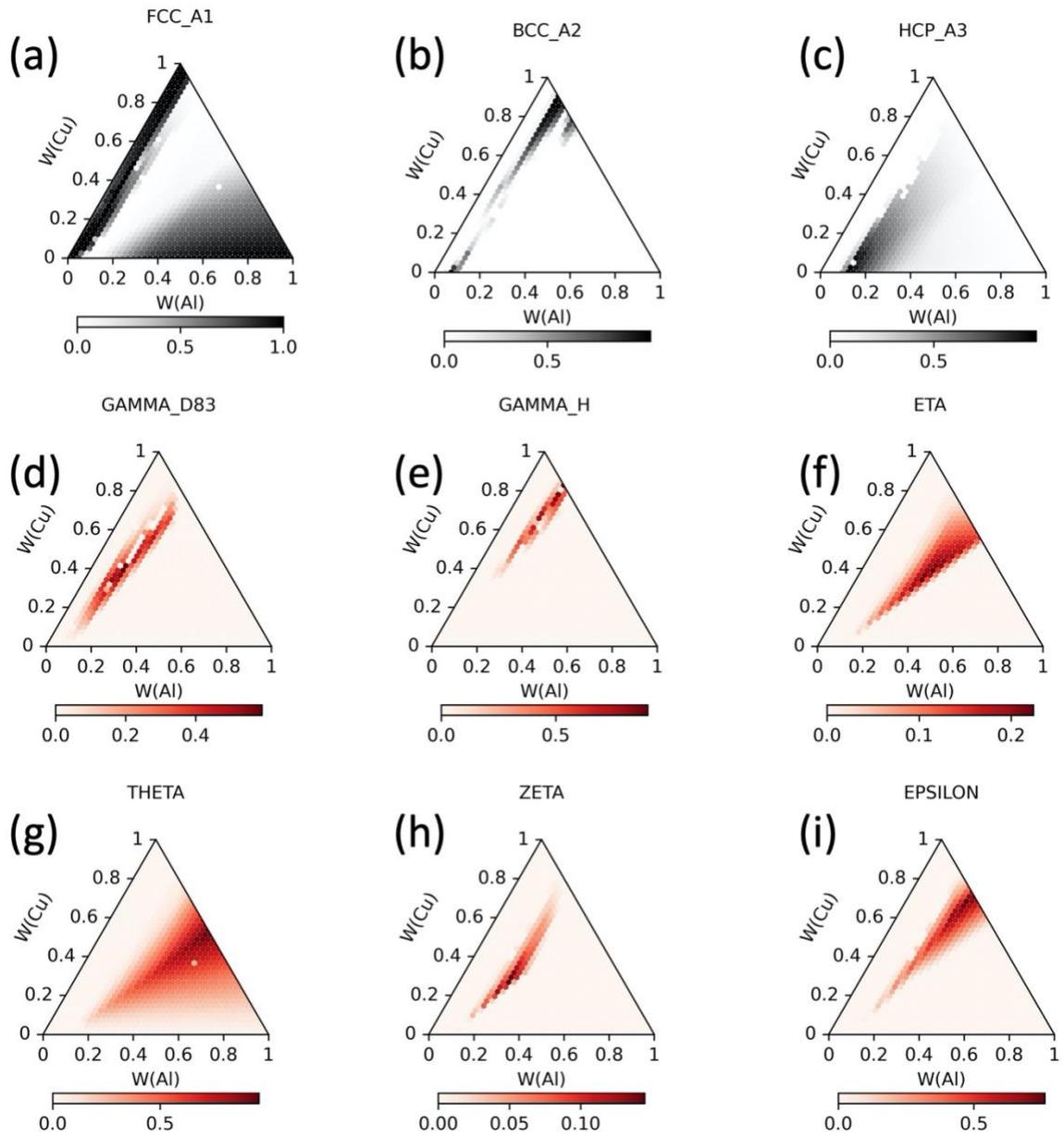

Figure 7 Heatmaps of predicted phases fraction in Ag-Al-Cu predicted from Scheil-Gulliver simulations.

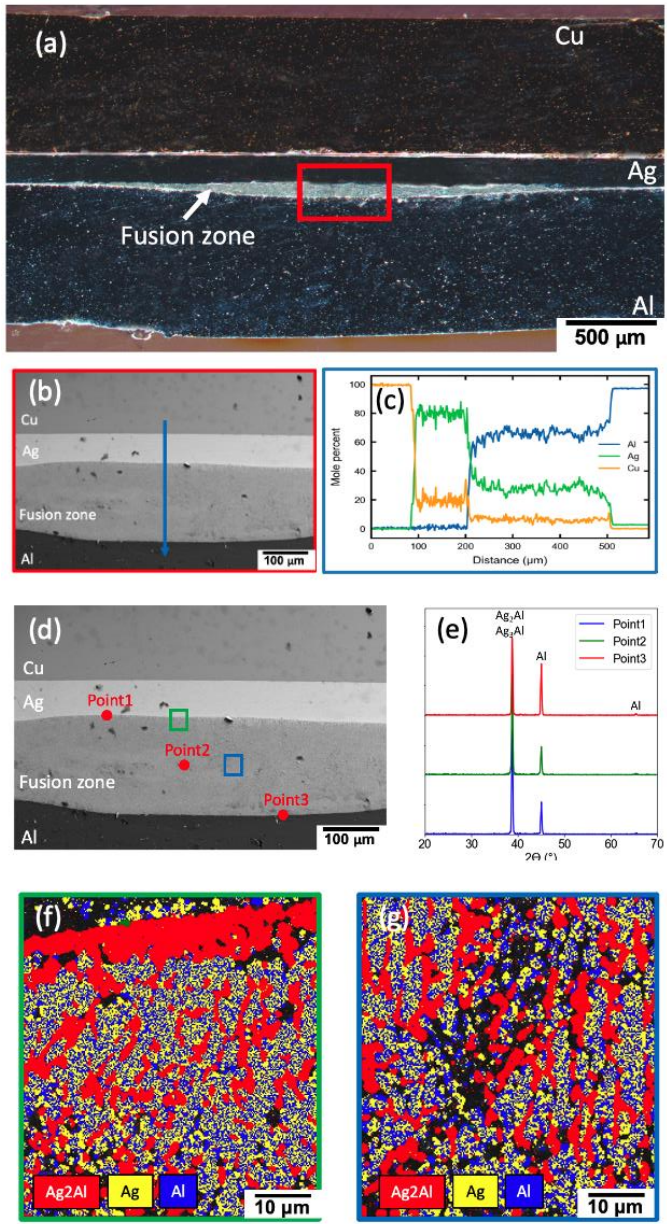

Figure 8 (a) Optical image, (b, c) EDS map, (d, e) XRD patterns, and (f,g) EBSD images of the RSW Al-Ag-Cu sample from the present work.

## Contents of Supplemental Material

Supplemental material includes two figures (Fig. S1 – Fig. S2 and a table as shown in the following pages.

Table. S1. The python code to run the feasibility map for Ag-Al-Cu with TDB from Witusiewicz et al. [49], output as Figures 4,7

```python
import os
import time
import matplotlib as mpl
import matplotlib.pyplot as plt
import numpy as np
import pandas as pd
from materialsmap.core.compositions import generateCompositions, createComposition
from materialsmap.ref_data import periodic_table, materials
from materialsmap.core.pycalphad_run import pycalphad_eq, pycalphad_scheil
from materialsmap.core.GenerateEqScript import createEqScript
from materialsmap.core.ReadEqResult import getEqdata
from materialsmap.core.GenerateScheilScript import createScheilScript
from materialsmap.core.ReadScheilResult import getScheilSolidPhase
from materialsmap.plot.FeasibilityMap import plotMaps

# Create Compositions
comps = ['Ag', 'Al', 'Cu']
eleAmountType = 'massFraction'
pressure = 101325
ngridpts = 41  # number of divisions along each dimension of the composition grid
TemperatureRange = (600, 2000, 10)  #(lower limit, upper limit, temperature step)
indep_comps = [comps[1], comps[2]]  # choose them automatically
maxNumSim = 250  # maximum number of simulations in each TCM file

# Equilibrium simulation settings
pressure = 101325
database = './Ag-Al-Cu_new4_as_published.TDB'  # <userDatabase>.TDB or TCFE8
eleAmountType = 'massFraction'  # Candidates: massFraction massPercent moleFraction molePercent
output_Eq = f'{TemperatureRange[0]}-{TemperatureRange[1]}-{TemperatureRange[2]}-{comps[0]}-{comps[1]}-{comps[2]}-Eq'

# Create folder in curent path to store simulation results
from datetime import datetime

current_dateTime = datetime.now()
if '.tdb' in database or '.TDB' in database:
    database_name = database.split('/')
    path = f'./Simulation/{datetime.now().strftime("%m-%d-%Y")}-{comps[0]}-{comps[1]}-{comps[2]}-database-{database_name[-1][:-4]}'
else:
    path = f'./Simulation/{datetime.now().strftime("%m-%d-%Y")}-{comps[0]}-{comps[1]}-{comps[2]}-database-{database}'
isExist = os.path.exists(path)
if not isExist:
    os.makedirs(path)
    print("The new directory is created!")
```

```python
# Save compostion results
compositions_list = generateCompositions(indep_comps, ngridpts)
Compositions, numPoint, comp, numSimultion = createComposition(indep_comps, comps,
compositions_list, materials, path)
settings = [TemperatureRange, numPoint, numSimultion, comp, comps, indep_comps,
os.path.abspath(database), pressure, eleAmountType]
np.save(f'{path}/setting.npy', settings)

# Running with PyCalphad
pycalphad_eq(path)
pycalphad_scheil(path, 2000)  # temperature to start scheil if not eq results

# Running with Thermo_Calc
# Create TCM files with path
createEqScript(path)
createScheilScript(path, 2000)  # temperature to start scheil if not eq results

# Open TCM files with Thermo_Calc

# Collect results from Thermo_Calc
getEqdata(path)
getScheilSolidPhase(path)

# Plot deleterious phase diagram and crack susceptibility map
plotMaps(path, 'pycalphad')
```

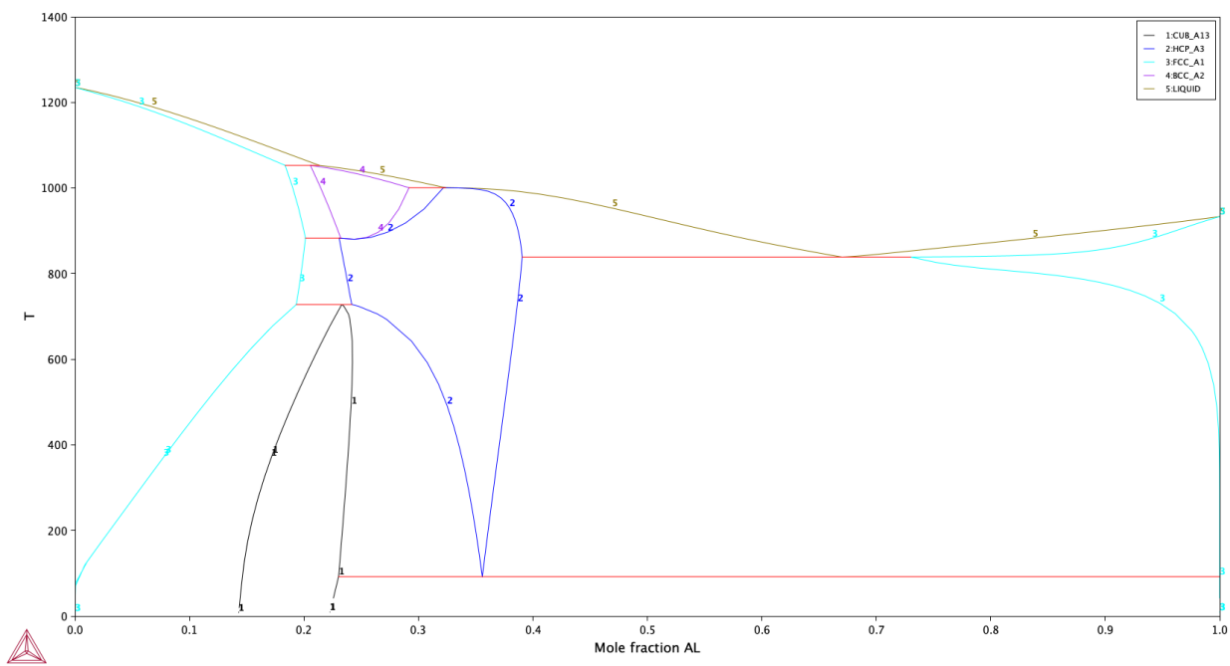

Figure S1. Al-Ag phase diagram from Witusiewicz et al. [49].